\documentclass[twocolumn,preprintnumbers, amssymb,amsmath,aps,floatfix,prl,nofootinbib,superscriptaddress,showpacs]{revtex4-1}

\usepackage{epsfig}
\usepackage{bm}
\usepackage{amssymb}
\usepackage{amsmath}
\usepackage{color}
\usepackage{subfigure}

\begin{document}

\title{Dijet Asymmetry in the Resummation Improved Perturbative QCD Approach}

\author{Lin Chen}

\affiliation{Key Laboratory of Quark and Lepton Physics (MOE) and Institute
of Particle Physics, Central China Normal University, Wuhan 430079, China}

\author{Guang-You Qin}
\author{Shu-Yi Wei}

\author{Bo-Wen Xiao}
\affiliation{Key Laboratory of Quark and Lepton Physics (MOE) and Institute
of Particle Physics, Central China Normal University, Wuhan 430079, China}

\author{Han-Zhong Zhang}
\affiliation{Key Laboratory of Quark and Lepton Physics (MOE) and Institute
of Particle Physics, Central China Normal University, Wuhan 430079, China}


\begin{abstract}

We develop the first systematic theoretical approach to dijet asymmetries in hadron-hadron collisions based on the perturbative QCD (pQCD) expansion and the Sudakov resummation formalism. We find that the pQCD calculation at next-to-leading order is indispensable to describe the experimental data, while the Sudakov resummation formalism is vital near the end points where the pQCD expansion fails to converge due to the appearance of large Sudakov logarithms. Utilizing our resummation improved pQCD approach, we obtain good agreement with the most up-to-date fully corrected ATLAS data on dijet asymmetry in $pp$ collisions. Combining with the BDMPS jet energy loss formalism, we extract the value of jet transport coefficient $\hat{q}_0 \sim 2$-$6~\textrm{GeV}^2/\textrm{fm}$ for the quark-gluon-plasma created in $PbPb$ collisions at 2.76A~TeV. This work paves the way for a more complete and deeper understanding of the properties of strongly-coupled QCD medium via the studies of dijet asymmetries in relativistic heavy-ion collisions. 

\end{abstract}
\maketitle


\textit{Introduction} --- Quantitative study of the strongly-coupled quark gluon plasma (QGP) created in heavy ion collisions at the Relativistic Heavy-Ion Collider (RHIC) and the Large Hadron collider (LHC) is one of the cutting-edge research topics in high energy nuclear physics. High energy jets traversing QGP medium undergo multiple scatterings with QGP which induces additional gluon radiation~\cite{Gyulassy:1993hr, Baier:1996kr, Baier:1996sk, Baier:1998kq, Zakharov:1996fv, Gyulassy:1999zd, Wiedemann:2000za, Arnold:2002ja, Wang:2001ifa}. This implies that jets in a strongly-coupled medium lose energies and receive additional transverse momenta due to the medium-jet interaction. Therefore, jet energy loss and transverse momentum broadening effects can be used as probes to study the so-called transport properties of QGP~\cite{Wang:1991xy, Majumder:2010qh, Qin:2015srf, Blaizot:2015lma}.
In the Baier-Dokshitzer-Mueller-Peigne-Schiff-Zakharov (BDMPS-Z) jet energy loss formalism~\cite{Baier:1996kr, Baier:1996sk, Baier:1998kq, Zakharov:1996fv}, the signature of both effects can be attributed to the so-called jet transport coefficient $\hat q$, which is defined as transverse momentum square transfer per unit length and reflects the density of QGP medium. By computing the nuclear modification factor for single hadron productions, JET collaboration has extracted the value of $\hat q$ according to RHIC and the LHC data \cite{Burke:2013yra}.  

In the era of the LHC, dijet process has become a vital tool for quantitatively studying the properties of quark-gluon plasma created in heavy-ion collisions. In particular, the difference between the distributions of dijet asymmetries in $PbPb$ and $pp$ collisions~\cite{Aad:2010bu, Chatrchyan:2011sx, Chatrchyan:2012nia} reveals that high energy jets tend to lose a significant amount of energy when traversing QGP medium created in $PbPb$ collisions~\cite{Qin:2010mn, CasalderreySolana:2010eh, Young:2011qx, He:2011pd, Lokhtin:2011qq, ColemanSmith:2012vr, Renk:2012cb, Zapp:2012ak, Ma:2013pha, Senzel:2013dta, Casalderrey-Solana:2014bpa, Ayala:2015jaa, Milhano:2015mng, Chang:2016gjp}. 
In the BDMPS-Z formalism, jet energy loss and transverse momentum broadening effects are two sides of the same coin and tightly related to each other. 
In our earlier studies~\cite{Mueller:2016gko, Mueller:2016xoc, Chen:2016vem}, we focused mostly on the studies of the transverse momentum broadening effect via back-to-back dijet (dihadron and hadron-jet) azimuthal angular correlation by using the so-called Sudakov resummation formalism~\cite{Banfi:2008qs, Mueller:2013wwa, Sun:2014gfa}. 

Previous theoretical studies of dijet asymmetries are mostly based on Monte Carlo simulations. In this paper, we present the first systematic investigation of this important observable based on the pQCD calculation supplemented with the Sudakov resummation at the end points. First, we find that, similar to event shape observables such as the thrust in $e^+e^-$ annihilations, the dijet asymmetry in proton-proton collisions can only reach certain range of the asymmetry distribution at given order in the pQCD expansions. Furthermore, at the end point where back-to-back dijet configurations dominate, the Sudakov resummation formalism is switched on since the pQCD expansion fails to converge. Using this combined theoretical framework, we nicely describe the up-to-date fully corrected dijet asymmetry data for $pp$ collisions measured by the ATLAS Collaboration~\cite{unfold, Perepelitsa:2016zbe} without fine-tuning.\footnote{Despite strong academic and phenomenological interests, all previous theoretical studies are based on the comparison with the uncorrected data~\cite{Aad:2010bu, Chatrchyan:2011sx, Chatrchyan:2012nia} which contains detector artifacts. This brings significant ambiguities in the quantitative studies of the energy loss effect. One should only compare our theoretical calculation of dijet asymmetries to the fully corrected data.} Using $pp$ results as the baseline, we add the energy loss effect \footnote{The contribution from transverse momentum broadening to dijet asymmetry is found to be negligible numerically as expected.}  on the dijet asymmetry in $PbPb$ collisions based on the BDMPS formalism, and find $\hat{q}_0 \sim 2-6 \textrm{GeV}^2/\textrm{fm}$ for $PbPb$ collisions at 2.76A~TeV.


\textit{Dijet asymmetry in $pp$ collisions} --- The subject of interest, namely, the dijet asymmetry is defined as $A_J \equiv \frac{p_{\perp 1}-p_{\perp 2}}{p_{\perp 1}+p_{\perp 2}}$, where $p_{\perp 1}$ and $p_{\perp 2}$ are the transverse momentum of the jet with the largest (leading jet) and second largest transverse energy (associated jet), respectively. 
One often defines a closely related variable $x_J\equiv \frac{p_{\perp 2}}{p_{\perp 1}}$ with the following one-to-one mapping $A_J=\frac{1-x_J}{1+x_J}$ or $x_J=\frac{1-A_J}{1+A_J}$. 
In fact, $A_J$ (or $x_J$) is a rather complex observable, since it not only involves the amplitude of the dijet momenta, but also implicitly encodes their angular distribution due to the underlying transverse momentum conservation. 
We should understand the distribution of $x_J$ from two different perspectives, namely, perturbative QCD expansion and Sudakov resummation. 
PQCD calculation is important in the small $x_J$ region, however it diverges in the threshold region $x_J \sim 1$, where the Sudakov formalism becomes extremely useful.

\begin{figure}[tbp]
\begin{center}
\includegraphics[width=1.0\linewidth]{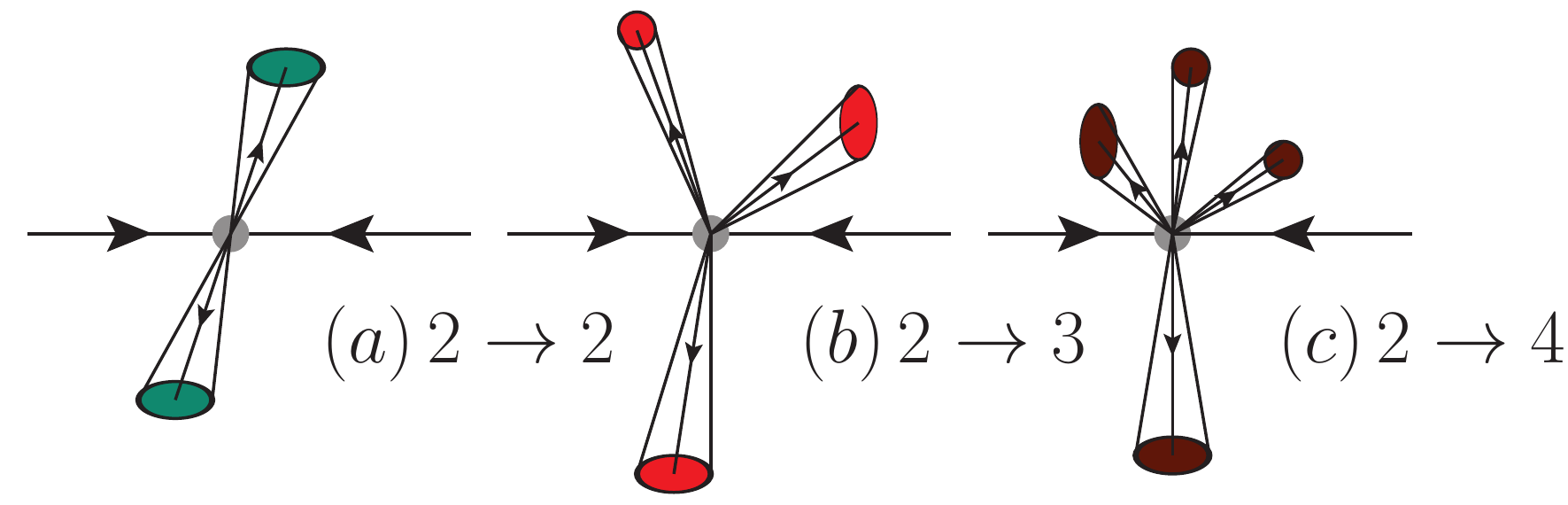}
\end{center}
\caption[*]{Illustration of (a) $2\to 2$ process with two jets in the final state; (b) $2\to 3$ process with three jets in the final state; (c) $2\to 4$ process with four jets in the final state.}
\label{jets}
\end{figure}

Let us begin with the perturbative QCD expansions as illustrated in Fig. \ref{jets}. In collinear framework, tree level $2\to 2$ process is the leading order (LO) for inclusive dijet total cross-section, but it is excluded for observables like $x_J$ (or $A_J$) and azimuthal angular correlation since this process gives trivial results such as $x_J=1$, $A_J = 0$ and $\Delta\phi = \pi$, respectively, simply due to transverse momentum conservation. 
Therefore, as far as the dijet asymmetry and angular correlation are concerned, the $2\to 3$ and $2\to 4$ processes as shown in (b) and (c), together with the corresponding virtual graphs,  correspond to the LO and NLO contributions, respectively. Based on Ref.~\cite{Nagy:2001fj, Nagy:2003tz}, which allows one to calculate dijet cross sections at NLO accuracy, we have numerically calculated $x_J$ distributions and the dijet angular correlation up to NLO for $pp$ collisions. 

Before presenting our numerical results, we first derive a very interesting lower bound for $x_J$ (upper bound for $A_J$) for a given $2\to n$ event. 
Assuming the following three conditions:
\begin{enumerate}
\item Total transverse momentum is conserved;
\item Experimental detectors are ideal with $4\pi$ coverage, which means there are no missing jets;
\item $p_{\perp 1}$ and $p_{\perp 2}$ are the momenta of jets with two largest transverse momenta,
\end{enumerate}
one can prove that  for $2\to n$ processes, the largest $A_J$ value that can be reached is $\frac{n-2}{n}$, which corresponds to the lower bound for $x_J=\frac{1}{n-1}$. 
Let us denote $p_{\perp i}$ with $i=1, 2, \cdots, n$ as the transverse momenta of $n$ final state jets with $p_{\perp 1}\geq p_{\perp 2}\geq \cdots \geq p_{\perp n}$.  
Using the above setup together with the transverse momentum conservation $\vec p_{\perp 1} + \vec p_{\perp 2} +\cdots + \vec p_{\perp n}=0$, we have:
\begin{eqnarray}
p_{\perp 1} &=& |\vec p_{\perp 1}| = | \vec p_{\perp 2} +\cdots + \vec p_{\perp n}| \notag \\
& \leq &  | \vec p_{\perp 2}| +\cdots + |\vec p_{\perp n}| \leq (n-1) p_{\perp 2} ,
\end{eqnarray}
therefore, $x_J \geq \frac{1}{n-1}$ and $A_J \leq \frac{n-2}{n}$ for $2 \to n$ processes.

\begin{figure}[tbp]
\begin{center}
\includegraphics[width=0.86\linewidth]{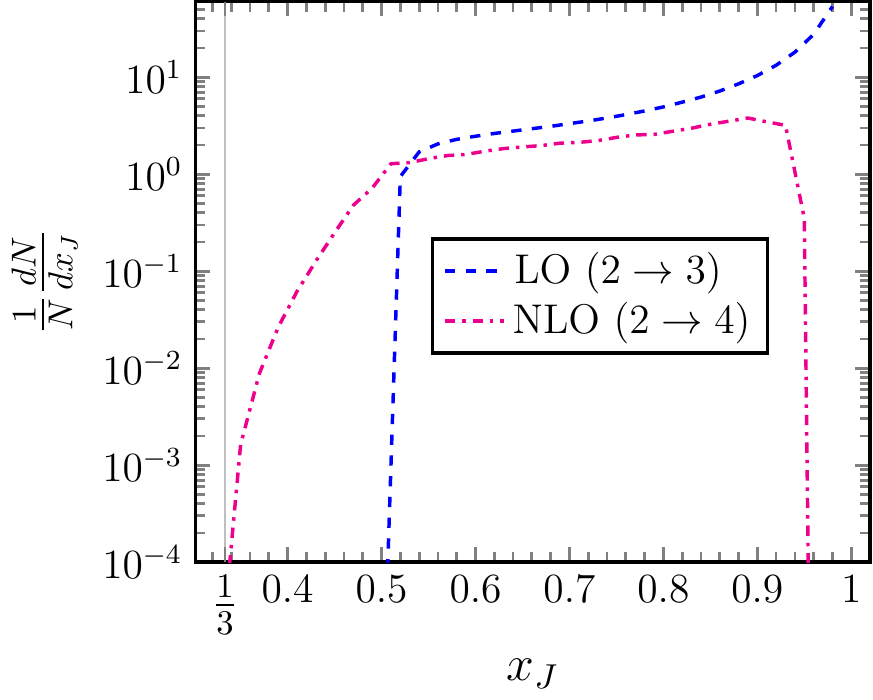}
\end{center}
\caption[*]{Normalized LO and NLO contributions to dijet $x_J$ distributions computed for $pp$ collisions at $2.76$~TeV with $p_{\perp 1} >100~\textrm{GeV}$ and $p_{\perp 2}  >50~\textrm{GeV}$ without rapidity cut.}
\label{aj-xj}
\end{figure}

The above lower bound for $x_J$ has interesting and substantial consequence on the perturbative calculation. For example, at LO which starts at $\mathcal{O}(\alpha_s^3)$, the smallest value that one can reach is approximately $\frac{1}{2}$ as shown in Fig.~\ref{aj-xj}, if the rapidity coverage is large enough which guarantees that missing jets are very rare. In the range where $\frac{1}{3} < x_J <\frac{1}{2}$, the NLO [$\mathcal{O}(\alpha_s^4)$] result becomes the dominant contribution. In the region where $x_J < \frac{1}{3}$, one has to rely on higher order expansions in order to get non-vanishing contributions. 
Following Refs.~\cite{Abazov:2004hm, Khachatryan:2011zj}, we set the normalizations in the perturbative calculations as $\frac{1}{\sigma_{\textrm{LO}}} \frac{d\sigma_{\textrm{LO}}}{dA_J}$ and $\frac{1}{\sigma_{\textrm{NLO}}} \frac{d\sigma_{\textrm{NLO}}}{dA_J}$ for the LO and NLO calculation, respectively. 

Another interesting result in Fig.~\ref{aj-xj} is that the perturbative expansions fails to converge around $x_J \sim 1$, where the LO contribution always tends to become large and positive, while the NLO contribution always turns large and negative. It is straightforward to trace the origin of the divergence of pQCD expansion back to the Sudakov-type large logarithms when the measured dijets are back-to-back. This can be clearly seen by studying the dijet angular correlations in $pp$ collisions using perturbative calculation and the Sudakov resummation formalism. 

\begin{figure}[tbp]
\begin{center}
\includegraphics[width=0.86\linewidth]{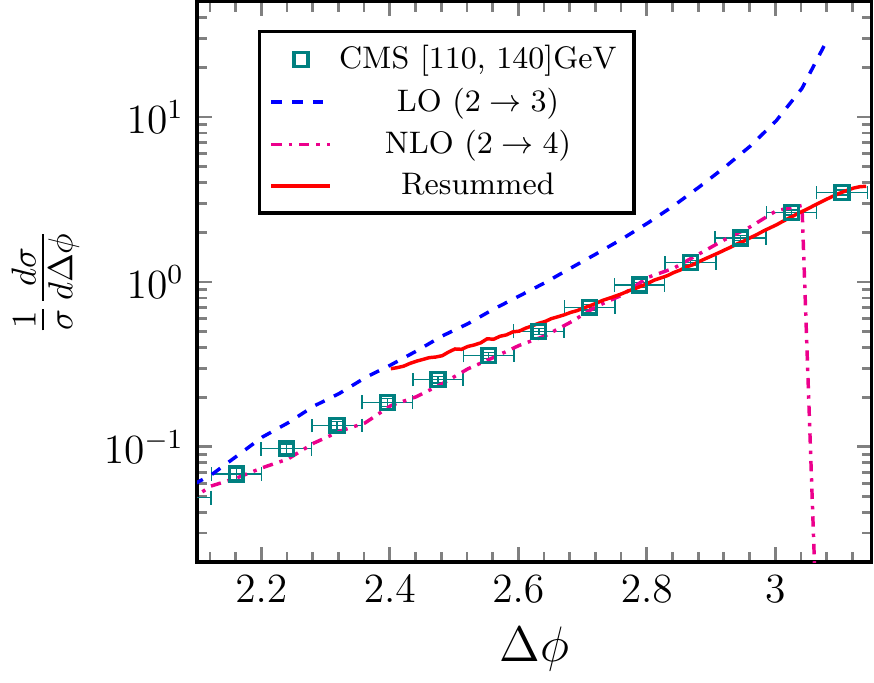} 
\end{center}
\caption[*]{The comparison between the CMS data~\cite{Khachatryan:2011zj} and the numerical calculation for the azimuthal angular correlation of dijets in $pp$ collisions at $\sqrt{s} =7~\textrm{TeV}$ in the range $110< p_{\perp 1} < 140~\textrm{GeV}$ and $p_{\perp 2} > 30~\textrm{GeV}$ with rapidity cut $|y|<1.1$, $R=0.5$, $\mu=2p_{\perp 1}$ and $\Delta \phi \equiv |\phi_1 -\phi_2|$.}
\label{correlation}
\end{figure}

In Fig.~\ref{correlation}, we plot the dijet angular correlation using at LO ($2\to 3$ real process only) and NLO (real and virtual $2\to 3$ graphs together with real $2\to 4$ contributions) as well as the Sudakov resummation formalism. 
The LO result can qualitatively describe the shape of the large angle deflection data when $\Delta \phi$ is far from $\pi$. The normalization of LO result is a bit off, since it is sensitive to the choice of the factorization scale $\mu$. In addition, it receives very large and positive logarithmic corrections [e.g., $\mathcal{O}(\alpha_s\ln^2 \frac{P_\perp^2}{q_\perp^2})$], with $\vec P_\perp\simeq \vec p_{\perp 1}$ and $\vec q_\perp = \vec p_{\perp 1}+\vec p_{\perp 2}$, around $\Delta \phi \sim \pi$. The NLO result, which is less sensitive to $\mu$, improves the LO calculation and matches the data much better, as expected. However, the NLO result suffers large but negative logarithmic corrections [e.g., $\mathcal{O}(\alpha_s^2\ln^4 \frac{P_\perp^2}{q_\perp^2})$] around $\Delta \phi \sim \pi$. In contrast, the Sudakov resummation formalism~\cite{Mueller:2013wwa, Sun:2014gfa} precisely captures this type of oscillating feature and resums the alternating sign series of large logarithmic corrections. 
The resummed result yields the correct description of the data around $\Delta \phi \sim \pi$. 
We should emphasize that the Sudakov formalism is only valid when $\Delta \phi$ is not far from $\pi$, since it performs the all-order resummation of the large logarithmic corrections [e.g., $\sum_{k=0}^{\infty} (-1)^k\mathcal{O}(\left(\alpha_s \ln^{2} \frac{P_\perp^2}{q_\perp^2}\right)^k)$], but neglects most of finite corrections. In practice, based on the Born contribution from $2\to 2$ process, the Sudakov resummation is carried out in the coordinate space in order to preserve transverse momentum for arbitrary number of soft gluon emissions.

The result in Fig.~\ref{correlation} clearly shows that the NLO calculation can reliably describe the dijet physics when $\Delta \phi$ is not too close to $\pi$, while Sudakov formalism is essential in order to describe the back-to-back dijet configurations for which naive pQCD expansion fails to converge in general. 
The same argument applies to Fig.~\ref{aj-xj}: the divergent behavior of the LO and NLO results also originates from the back-to-back dijet configurations, since typical back-to-back dijet events give $x_J\sim 1$. Therefore, in this particular region, one should replace pQCD expansion with the resummed calculation. 

To compute the $x_J$ and $A_J$ distribution and compare with experimental data, here we propose to use the following Sudakov resummation improved NLO formula:
\begin{eqnarray}
 \left.\frac{1}{\sigma} \frac{d\sigma}{dx_J}\right|_{\textrm{improved}} &=& \left.\frac{1}{\sigma_{\textrm{NLO}}} \frac{d\sigma_{\textrm{NLO}}}{dx_J}\right|_{\Delta \phi <\phi_{\textrm{m}}}
\nonumber\\&&
+\left.\frac{1}{\sigma_{\textrm{Sudakov}}} \frac{d\sigma_{\textrm{Sudakov}}}{dx_J}\right|_{\phi_{\textrm{m} < \Delta \phi <\pi}}, \label{improved}
 \end{eqnarray}
where $\phi_{\textrm{m}}$ is the matching point which separates the phase space in the dijet angular distribution between the Sudakov formalism and NLO calculation.
Here we set $\phi_m$ as the maximum point of the NLO curve, i.e., around $\Delta\phi =3$ (see Fig. \ref{correlation}), and we have checked that the final $x_J$ distribution does not change much with different choices of $\phi_m$. 
A few comments follow the above formula. 
First, $\left.\frac{1}{\sigma} \frac{d\sigma}{dx_J}\right|_{\textrm{improved}}$ is, in principle, automatically normalized to unity, since $\frac{1}{\sigma_{\textrm{NLO}}} \frac{d\sigma_{\textrm{NLO}}}{dx_J}$ and $\frac{1}{\sigma_{\textrm{Sudakov}}} \frac{d\sigma_{\textrm{Sudakov}}}{dx_J}$ have been properly normalized, respectively. Second, we believe there is no double-counting when combining two different formalisms here. As far as the pQCD expansion is concerned, we have been able to compute the complete cross section at the NLO accuracy. In addition, by matching with the Sudakov formalism, we have included all the large logarithms such as $\left[\alpha_s\ln^2 \frac{P_\perp^2}{q_\perp^2}\right]^k$ at next-to-leading logarithmic (NLL) level. 
The contributions that have been neglected are of $\mathcal{O}(\alpha_s^5)$ and those beyond NLL approximation. Last but not least, there is no free parameter in this calculation, since the results are not very sensitive to choice of the factorization scale $\mu^2$ or the matching point $\phi_m$. 

\begin{figure}[tbp]
\begin{center}
\includegraphics[width=0.73\linewidth]{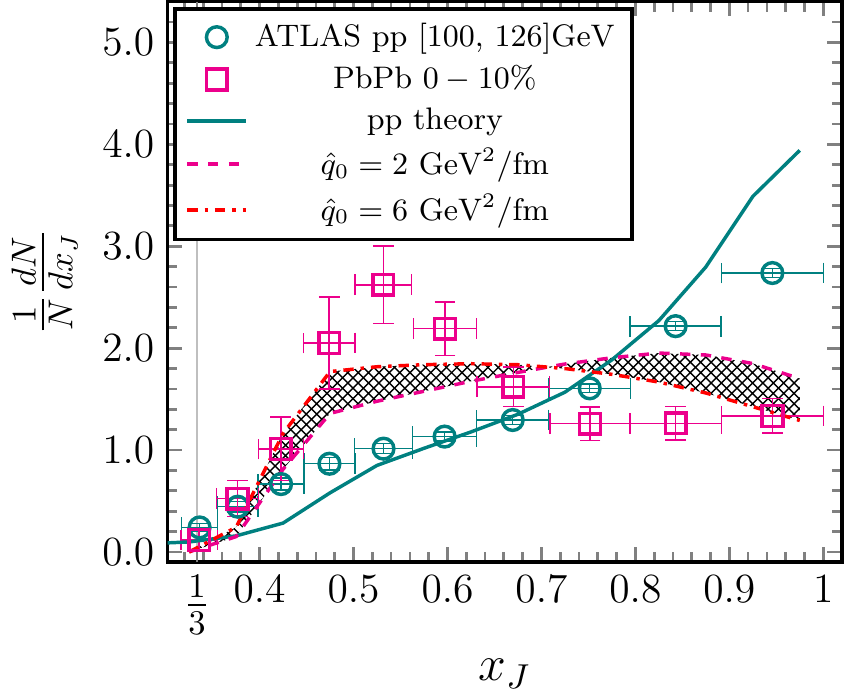}  
\includegraphics[width=0.73\linewidth]{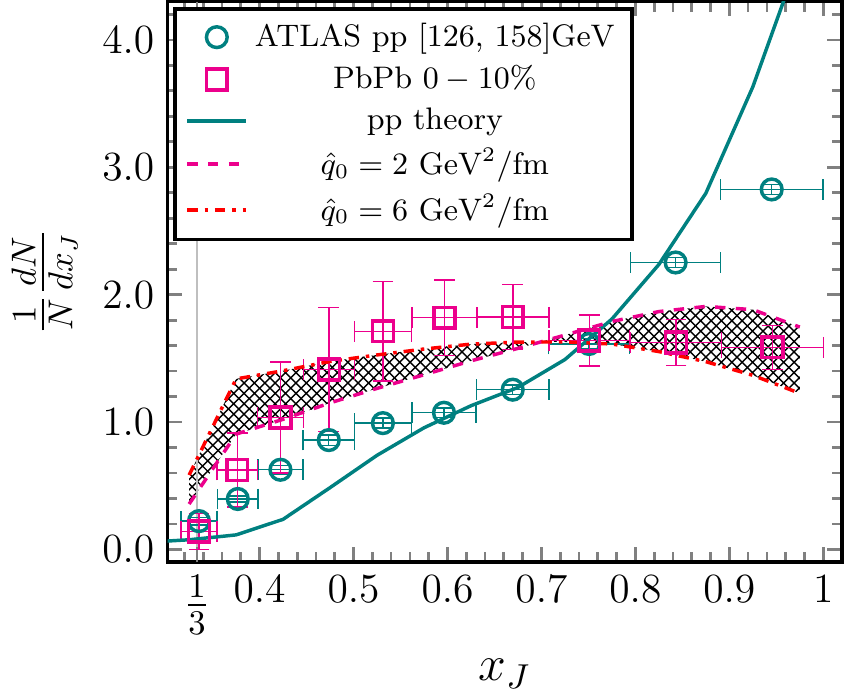}  
\includegraphics[width=0.73\linewidth]{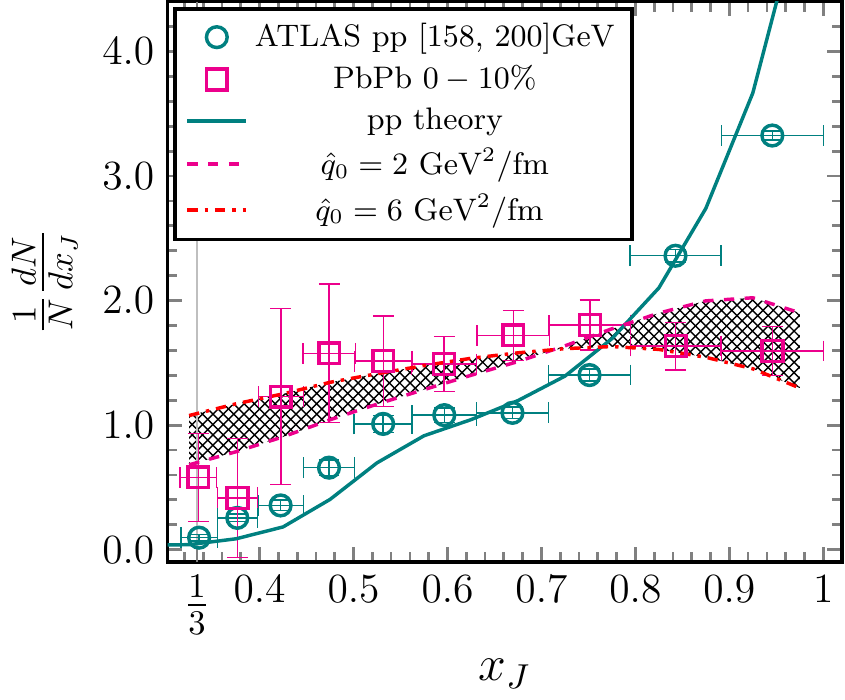}  
\includegraphics[width=0.73\linewidth]{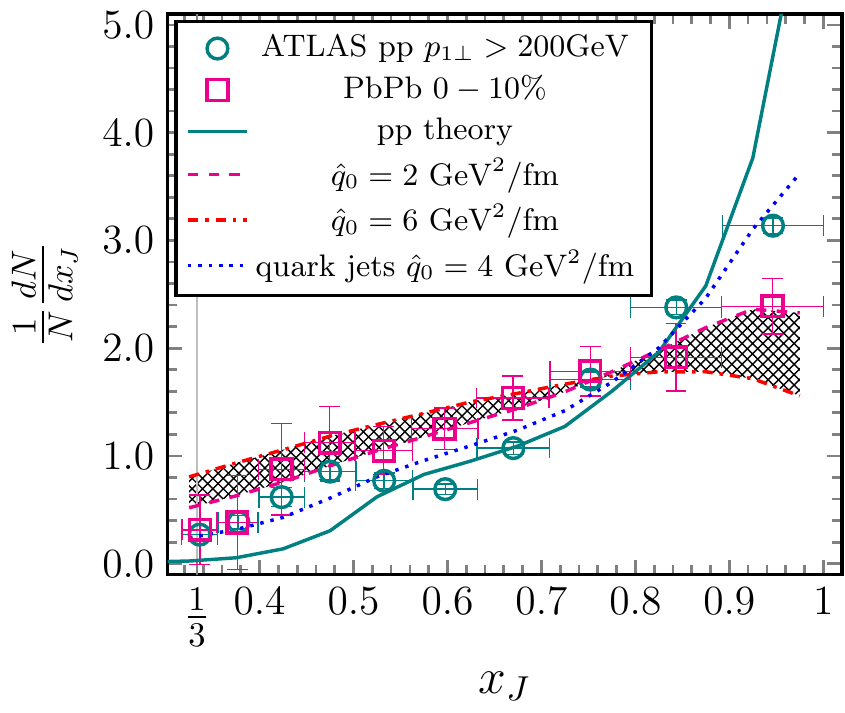}
\end{center}
\caption[*]{Normalized dijet $x_J$ distributions computed for $pp$ and $PbPb$ collisions at $2.76$A~TeV with $p_{\perp 1} > p_{\perp 2}  >25~\textrm{GeV}$, $R=0.4$ and rapidity cut $|y|<2.1$ compared with the fully corrected data taken from Ref.~\cite{unfold, Perepelitsa:2016zbe}. The theoretical bands are obtained by varying $\hat{q}_0$ from $2~\textrm{GeV}^2/\textrm{fm}$ to $6~\textrm{GeV}^2/\textrm{fm}$. }
\label{atlasxj}
\end{figure}

In Fig.~\ref{atlasxj}, we show the dijet asymmetry $x_J$ distributions (in $pp$ collisions) from our Sudakov resummation improved NLO framework, namely Eq.~(\ref{improved}), and compare to the fully corrected data from the ATLAS collaboration~\cite{unfold, Perepelitsa:2016zbe}, for four different $p_\perp$ ranges.
We emphasize that the dijet asymmetry data published in Ref.~\cite{Aad:2010bu, Chatrchyan:2011sx, Chatrchyan:2012nia} contain the contributions from experimental artifacts, such as the detector response and local underlying event fluctuations, which have to be removed through the so-called unfolding process before comparing to our theoretical calculations. 
We can see that dijet asymmetry distribution in the large $x_J$ region can be nicely described by our resummation improved pQCD approach. 
In the small $x_J$ region, although the shape agrees with the data, the magnitude is a bit smaller, which is mainly due to the lower bound $x_J=\frac{1}{3}$. 
The NLO contribution has to vanish near $x_J=\frac{1}{3}$ and the Sudakov contribution is also diminishing very fast there. 
This implies that higher order (e.g., next-to-next-to-leading-order) contributions could become important in the small $x_J$ region in order to fully describe the dijet asymmetry data. We have checked that this small difference between our theoretical result and $pp$ data affect very little on the extraction of jet transport coefficient of QGP created in $PbPb$ collisions in the following section, when we use $x_J$ distributions in $pp$ collisions as the baseline.

\textit{Dijet asymmetry in $PbPb$ collisions} --- In order to study the energy loss effect on dijet asymmetry distribution in heavy-ion collisions, and to quantitatively extract the value of jet transport coefficient, we embed the above theoretical framework for the dijet asymmetry calculation in a realistic modelling of the collision geometry and the space-time evolution of the QGP medium simulated via the OSU (2+1)-dimensional viscous hydrodynamics code~\cite{Song:2007ux, Qiu:2011hf}. 
According to the locations of the hard collisions and the propagation directions of the produced jets, we can calculate the amount of energy loss experienced by each jet traversing the QGP medium. 
As a first step, we employ a simple energy loss distribution derived in the BDMPS formalism~\cite{Baier:2001yt} to calculate the energy loss for dijets. 
In the limit of high energy jet and small fractional energy loss (i.e., $\epsilon/p_\perp \ll 1$), the energy loss $\epsilon$ distribution due to multiple soft gluon emissions can be written as: 
\begin{equation}
D(\epsilon) =\alpha \sqrt{\frac{\omega_c}{2\epsilon}}\exp \left(-\frac{\pi\alpha^2 \omega_c}{2\epsilon}\right),
\end{equation}
where $\omega_c \equiv \frac{1}{2} \hat{q}_R L^2$ and $\alpha \equiv \frac{2\alpha_s (\mu_r^2) C_R}{\pi}$ with $C_R=C_F (N_C)$ for quark (gluon) jets. For typical value of $\mu_r^2\sim \hat q L \sim 10~\textrm{GeV}^2$, the strong coupling $\alpha_s \simeq 0.2$. 
Here we parameterize the quark jet transport coefficient as: $\hat{q}=\hat{q}_0T^3/T_0^3$, with $T_0=481~\textrm{MeV}$ for $PbPb$ collisions at $2.76$A~TeV at the LHC. 
To further simplify the calculation, we assume all produced jets are gluon jets since they are the dominant part of measured jets at the LHC for the trigger jet $p_{1\perp} < 200~\textrm{GeV}$. 
Also to compare our result with the data in Ref.~\cite{unfold}, we have increased the lower cut for the associated jet $p_{2\perp} >45~\textrm{GeV}$ for all four $p_\perp$ bins. 
This is motivated by the fact that lower energy jets measured in $PbPb$ collisions suffer from large corrections (as mentioned in Ref.~\cite{unfold}, the correction for $30$~GeV jets can be as large as 40\%), while the unfolding procedure tends to move the associated jet $p_{2\perp}$ towards the leading jet $p_{1\perp}$. 

As shown in Fig.~\ref{atlasxj}, $\hat{q}_0$ $\sim$ $2$-$6~\textrm{GeV}^2/\textrm{fm}$ can reasonably describe the data in central $PbPb$ collisions at the LHC, for three relatively low $p_\perp$ bins. For the highest $p_\perp$ bin ($p_{1\perp} > 200~\textrm{GeV}$), the data favors small value of $\hat{q}_0 \sim 2~\textrm{GeV}^2/\textrm{fm}$. 
We believe that this is simply due to the increase of the quark jet fraction with increasing jet energies (see the dotted blue curve which is obtained by assuming all jets are quark jets with $\hat{q}_0=4~\textrm{GeV}^2/\textrm{fm}$).
The above extracted value of $\hat{q}_0$ amounts to a typical energy loss of about $20$-$30~\textrm{GeV}$ for gluon jets, and this value is consistent with our previous finding via dihadron and hadron-jet angular correlation calculation~\cite{Chen:2016vem}. Assuming the scaling law $\hat{q}=\hat{q}_0T^3/T_0^3$, one can find that $\hat{q}$ $\sim$ $0.3$-$0.8~\textrm{GeV}^2/\textrm{fm}$ at $T=250~\textrm{MeV}$, which is in agreement with the original BDMPS estimate~\cite{Baier:1996sk}. The data also seems to suggest that the energy loss of jets at the LHC has weak or little dependence on jet energies.


\textit{Conclusion} --- In summary, we have developed the Sudakov resummation improved pQCD formalism to compute the dijet asymmetry distributions for both $pp$ and $PbPb$ collisions at the LHC energies.
Our $pp$ calculation can well describe the fully corrected dijet asymmetry data from the ATLAS Collaboration. 
Combining with the BDMPS jet energy loss formalism, we have obtained the jet transport parameter $\hat{q}_0 \sim 2$-$6~\textrm{GeV}^2/\textrm{fm}$ by comparing with the dijet asymmetry data in central $PbPb$ collisions at $2.76$A~TeV. 
This work serves as a benchmark calculation without much theoretical ambiguity for utilizing dijet asymmetry as a quantitative and precise tool to probe the jet-medium interaction in relativistic heavy-ion collisions.


\textit{Acknowledgment} --- We thank Al Mueller, Feng Yuan and Ben-Wei Zhang for stimulating discussions and interesting comments. 
This material is based upon the work supported by the Natural Science Foundation of China (NSFC) under Grant Nos.~11575070, 11375072 and 11435004, and the Major State Basic Research Development Program in China under Contract No. 2014CB845400. 


\bibliographystyle{h-physrev5}
\bibliography{refs_dijet}

\end{document}